\newcommand{\be}{\begin{equation}}
\newcommand{\ee}{\end{equation}}
\newcommand{\bea}{\begin{eqnarray}}
\newcommand{\eea}{\end{eqnarray}}

\documentstyle[12pt,cite]{article}

\newcommand{\PL}[3]{{Phys. Lett.}        {#1} {(19#2)} {#3}}

\newcommand{\NP}[3]{{Nucl. Phys.}        {#1} {(19#2)} {#3}}

\newcommand{\beq}{\begin{equation}}
\newcommand{\eeq}{\end{equation}}
\newcommand{\beqa}{\begin{eqnarray}}
\begin{document}
\renewcommand{\thefootnote}{\fnsymbol{footnote}}
\thispagestyle{empty}
\rightline{LNF-96/023 (IR)}
\rightline{hep-ph/9605374 \vspace{1cm}}
\begin{center}
{\bf PHYSICS CAHIERS -- No. 1}\footnote{Pacs No.: 10.} \vspace{1.5cm} \\
S. Bellucci${}^a$\footnote{E-mail: bellucci@lnf.infn.it}
                    \vspace{1cm} \\
${}^a$INFN-Laboratori Nazionali di Frascati, P.O.Box 13 I-00044 Frascati,
            Italy
\vspace{2.5cm} \\
{\bf Abstract} 
\end{center}
We begin herewith the editing of physics notes taken in the course
of Journal Club seminars at INFN-LNF in 1996.
The activity consists of informal talks about work in progress
and/or review of (more or less) recent physics results of interest to
our laboratory. In the section titles the name of the speakers appear,
together with the topics discussed in the seminars.
We plan to publish these notes twice a year.
\vspace{1.5cm} \\
\begin{center}
{\bf Table of contents} 
\end{center}
1 S. Bellucci: In-medium ${\bar K}$N scattering and chiral lagrangians\\
2 M. Greco: QCD jets at high p$_T$\\
3 G. Isidori: The problem of $R_b$: a phenomenological update\\
4 D. Babusci: DHG sum rule and the nucleon spin polarizability at LEGS\\
5 C. Forti: Underground muons, a tool to study the
cosmic ray composition\\
6 R. Baldini: Question marks in the nucleon time-like form factors\\
7 G. Pancheri: Eikonalized minijets cross-section in photon collisions
\vfill
\begin{center}
May 1996
\end{center}
\setcounter{page}0
\renewcommand{\thefootnote}{\arabic{footnote}}
\setcounter{footnote}0
\newpage
\section{S. Bellucci: In-medium ${\bar K}$N scattering and chiral
lagrangians, or the disappearence of $\Lambda$(1405) in heavy kaonic
atoms$^*$}

{\bf *} Also presented at 2$^{nd}$ DEAR Collaboration Meeting, LNF, 1-2
April 1996.

My message here is twofold: 
\begin{itemize}
\item there is a clean prediction of the chiral
symmetry effective lagrangian, stating that the K$^-$p scattering length
in a medium
strongly depends on the nuclear density, so that its real part changes sign
already at 1/8 of the normal nuclear density \cite{a1};
\item a not negative kaon scattering length on an isolated proton (kaonic
hydrogen) can be obtained only for values of the coupling constants
among the 6 coupled meson-baryon channels,
which are not compatible with SU(3) flavour symmetry.
These constants are calculated in \cite{a2}.
\end{itemize}
In what follows, I focus on the first point, which is a signal to
the experimenters \cite{a3}
of the importance of a measurement for atoms heavier than hydrogen
and deuterium. For the second point, see also \cite{a4}.

A recent study \cite{a1} obtains the following interesting results.
Nuclear matter modifies very strongly the low-energy K$^-$p
interaction. The attractive forces that produce the $\Lambda$(1405)
as a bound state are reduced by Pauli blocking. In medium the
$\Lambda$(1405) moves above the K$^-$p threshold at one-eighth
the normal nuclear matter density. The K$^-$p scattering length
depends strongly on the density. Its real part changes sign
at one-eighth the normal nuclear matter density. Correspondingly
the optical potential for kaonic atoms has an unconventional
$r$-behaviour. The presence of the $\Lambda$(1405) bound state
just 27 MeV below the K$^-$p threshold invalidates the low-density
theorem - stating that the optical potential goes linearly with
the density - at unusually low density values.

The microscopic understanding of the above features is based on the
low-energy QCD. A dynamical model of the $\Lambda$(1405) structure
as a bound state of ${\bar K}$ and N in the I=0 channel (and a resonance
in the $\Sigma\pi$ channel), based on the iteration of a pseudo-potential
to infinite order in a Lippmann-Schwinger equation and describing
successfully the data in the S=-1 strangeness sector, is modulated
to respect the SU(3) chiral symmetry and have, in the Born approximation
and up to terms of order O($q^2$) in the meson momentum, the same
s-wave scattering length as the effective chiral lagrangian
describing the low-energy meson-baryon interaction \cite{a2}.

The successes of the theory in describing the s-wave coupled channel
dynamics of the ${\bar K}$N and $\pi$-hyperon systems (i.e. the
$\Lambda$ binding energy, its width, and all available low-energy data 
of ${\bar K}$N, $\Sigma\pi$, $\Lambda\pi$ systems), persist in
describing how the nuclear matter affects the formation of the
${\bar K}$N bound state. At a small density value, i.e. far out in the
nuclear surface where the nuclear density distribution has some
overlap with the atomic K$^-$ wave function, the bound state disappears,
as the Pauli exclusion principle yields enough repulsion energy
to compensate the binding energy $E_{\Lambda}=-27$ MeV. The K$^-$p
amplitude varies rapidly with the density near threshold, hence the
effective scattering length in nuclear matter has a strong dependence
on the density. The corrections to the K$^-$p amplitude due to
the Fermi motion and the nucleon binding are also evaluated. They turn
out to be much less important (and mutually opposing) effects, in
comparison with the Pauli blocking of intermediate states \cite{a1}.

\section{M. Greco: QCD jets at high p$_T$}

In a recent CDF paper \cite{b1} the comparison between the data
- starting from a rather low p$_T$ ($\ge 15$ GeV)
and up to a maximum of 400 GeV -
and the complete O($\alpha_s^3$)
calculation \cite{b2,b3} of the inclusive jet cross-section
$$\frac{d\sigma^{jet}}{d^3p_T},$$
at large p$_T$ with the rapidity ranging in the interval
0.1$\le |\eta |\le$0.7,
is carried out. The agreement is quite good for about ten orders
of magnitude. However it
appears that a small discrepancy is present at large p$_T$,
i.e. the data suggest a
departure from the O($\alpha_s^3$)
prediction for $E_T \ge$ 200-250 GeV. One must bear in mind that
the D0 collaboration reported very recently new data showing
no deviation effect in the same range. Due to the
large errors in the high  p$_T$ region, the latter are in agreement
with both the QCD calculation and the CDF data. The discrepancy
has been indicated as a possible signal for quark compositeness.
A composite scale $\Lambda =$1-2 TeV has been estimated
by means of an effective four-fermion interaction
$$\frac{1}{\Lambda^2}{({\bar q}\Gamma q)}^2 .$$ However, before
drawing any definite conclusion, it is important to evaluate
carefully the theoretical uncertainties in the QCD prediction,
particularly at large $E_T$.

The data in the region of discrepancy are in the large $x$ region
($x\ge 0.6$) where the structure functions, in particular
the gluon one, are not very well measured. Indeed the CTEQ collaboration
has tried to change
the g structure functions and checked the corresponding effect
\cite{b4}. Generally the theoretical uncertainties in the complete
O($\alpha_s^3$) calculation due to the changes
in the scales and the structure functions are of order 20-30$\%$.
In the estimate of the full theoretical prediction
one needs however to take into account also the corrections coming
from higher orders, which could become relevant near the boundary of
the phase space ($p_T \simeq \sqrt{s}$/2).
Indeed, in the large $x$ region
the QCD expansion parameter is ${\biggr[ \alpha_s ln(1-x)\biggr] }^n$,
rather than $\alpha_s$. Hence,
when $x$ is close to 1 and correspondingly
$\biggr[ \alpha_s ln(1-x)\biggr] =$
O(1), one needs to resum all these terms, i.e. a finite order calculation
is not enough. The Dokshitzer-Gribov-Lipatov-Altarelli-Parisi (DGLAP)
equation gets modified \cite{b5}. All those terms which diverge for
$x\to$1 are related
to soft and collinear gluon radiation.
In Drell-Yan processes this correction has been calculated and
the effect of $x\to$1 is important, but it has not
been measured so far. The CDF data would eventually
yield the first opportunity to measure this effect.
The claim \cite{b6}
is that the corrections increase the value of the inclusive jet
cross-section at large $p_T$, i.e. they go in the right direction.

If the corrections turn out to be large, then one needs also to
correct for $x\to 1$ the structure function code used in analyzing the
DIS data. Of course
before making any claim about possible evidence
for the preonic structure of the quarks, one must include the
effect of these higher order corrections.

\section{G. Isidori: The problem of $R_b$: a phenomenological update}

\paragraph{1.} Electroweak precision tests performed at LEP and 
SLC have confirmed the Standard Model (SM) predictions with great 
accuracy. Among several observables which have been measured, 
only the ratio $R_b = \Gamma(Z\to b\bar{b})/\Gamma(Z\to \hbox{hadrons})$  
shows a departure form the SM prediction larger than three standard 
deviations. In particular, the most recent determination of 
$R_b$ obtained by combining the four LEP--experiments, 
$R_b^{exp}=0.2219\pm 0.0017$ \cite{LEP}, is $3.5\sigma$ far form the SM 
prediction: $R_b^{SM}=0.2157\pm 0.0001$.

\paragraph{2.} Within the SM, due to non--decoupling 
effects induced by the top quark, the $Z\to b\bar{b}$ vertex 
receives non--universal corrections  \cite{Bernabeu,Barbieri}.
The triangular diagrams $Z\to t\bar{t} \to  q_d\bar{q_d}$ ($W$-exchange)
and $Z\to W^+ W^- \to  q_d\bar{q_d} $ ($t$-exchange) are completely 
negligible for $q_d=d$ (due to small CKM matrix elements) whereas they
are relevant for  $q_d=b$. Interestingly, these effects are proportional to 
the Yukawa coupling of the top (this is the reason why they do not 
decouple as $m_t \to \infty$) i.e. they
are related to the symmetry--breaking breaking sector of the Model. 

The leading correction induced by top loops can be written as a modification of
the universal down--type coupling constants of the $b$ quark
with the $Z$ boson: $g_V^b = g_V^d +\tau/2$ and  $g_A^b = g_A^d +\tau/2$,
where $\tau=(g_t/4\pi)^2$ and $g_t=m_t/(2\sqrt{2}G_F)^{-1}\simeq 1$ is the 
Yukawa coupling of the top. Since both $g_V^d$ and $g_A^b$ are negative 
the resulting effect is a {\it decrease} of $\Gamma(Z\to b\bar{b})$
with respect to $\Gamma(Z\to d\bar{d})$. 

Besides the modification of $g_{V,A}^b$ induced by top exchanges,
there are other non--universal corrections which affect
$\Gamma(Z\to d\bar{d})$ \cite{Hollik}: 
phase space modifications due to $m_b\not=0$,
of order $O(m_b^2/M_Z^2)$; 
strong corrections calculated with $m_b\not=0$, of order
$O(\alpha_s m_b^2/M_Z^2)$; 
strong corrections to one--loop top--exchange diagrams
$\sim O(\alpha_s g_t^2/16\pi^2)$. All the corrections 
have been calculated (up to two loop in many cases
\cite{Barbieri,Hollik}) and the corresponding uncertainties
are negligible ($\delta R_b^{SM}\simeq 10^{-4}$ is a very
conservative estimate).

\paragraph{3.} Assuming that the discrepancy of $R_b$ is generated by 
non--SM physics:  $R_b^{exp} = R_b^{SM} (1+\delta^{non-SM})$,  
then also the determination of $\alpha_s(M_Z)$ 
performed at LEP via the ratio 
$R_h = \Gamma(Z\to \hbox{hadrons})/\Gamma(Z\to \mu^+\mu^-)$  
has to be modified. The value of $\alpha_s(M_Z)$ extracted by $R_h$
introducing $\delta^{non-SM}$ is lower than the 
uncorrected value and is in better 
agreement with the low energy determinations
(from $\sim 2\sigma$ above the value shift to $\sim 1\sigma$  below
\cite{Langaker}). Though not 
very significant form the statistical point of view, 
this result enforces the hypothesis of new physics in
$\Gamma(Z\to b\bar{b})$. On the other hand, playing the same game
with $R_c$, whose experimental value is about $2\sigma$
below the SM value, the resulting value of $\alpha_s$ is completely 
inconsistent with the low energy determinations. 

\paragraph{4.} New physics sources in the process 
$Z\to b\bar{b}$ can be generally divided in two classes: 
loop  and tree--level effects.  Let us start to analyze the former.

The contribution generated by a fermion $(F)$, with 
the relative Yukawa scalar field $(A)$,
to the triangular diagram $Z\to b\bar{b}$
can be calculated in a model independent way (imposing 
only the conservation of charge and weak--isospin) \cite{bo}.
The sign of the correction thus obtained depends crucially from the 
weak--isospin assignment of $F$ and can be applied to several
interesting cases:
\begin{itemize}
\item{\bf Two Higgs doublets.} In this case $F= t$ and $A = H^\pm$
(the new physical--charged--Higgs field). 
Like in the SM, the correction is negative and there is no possibility 
to improve the agreement with the data. 
\item{\bf Fourth Generation.} In this case $F= t'$ (the new up--type quark)
and $A=\phi^\pm$ (the
SM unphysical--charged--Higgs field, which in the unitary gauge appears as  
the $W$ longitudinal degree of freedom). Also in this case 
the correction is negative.
\item{\bf MSSM.} In this case there are three separate contributions:
top and charged--Higgs, charginos and stop, neutralinos and sbottom.
The first one has a negative sign (as in the
Two Higgs doublet case) whereas the second and the third one
can have a positive sign. 
For high ($\sim 1$ TeV) and almost degenerate values of SUSY particle masses
the three contribution cancel each other. Only for light stop and 
charginos (with small $\tan\beta$) or light sbottom and neutralinos
(with very large $\tan\beta$) there is a chance to improve the 
agreement with the data. However, a recent correlated analysis of MSSM
parameters (including new LEP data, $b \to s\gamma$ and 
Tevatron results) shows that is impossible to decrease to less than
$2 \sigma $ the discrepancy \cite{Ellis}. Furthermore, if this was the case, 
then light SUSY particles should be in the  LEP200 range.
\end{itemize}

For what concerns tree--level effects, recently it has been shown that
a leptophobic $Z'$ \cite{Alta}, universally coupled to up-type and 
down-type quarks, not only can generate the right correction to $R_b$
but can also improve the agreement with  CDF data of the inclusive 
jet cross section at high $p_T$ ($p_T \geq 200$ GeV) \cite{b1}.
The mass of the   
$Z'$ is estimated to be in the TeV range. 

\paragraph{5.} To conclude, we can say that there is no clear solution to the
problem of $R_b$ yet. The possibility of new heavy fermions with 
unconventional  weak isospin assignment or the leptophobic--$Z'$  hypothesis
point in the right direction but are still {\it ad hoc} solutions. 
On the other hand, the possibility of a statistical fluctuation in the 
experimental data  is far from being excluded. 
\newpage
\section{D. Babusci: DHG sum rule and the nucleon spin polarizability at LEGS}

Energy weighted integrals of the difference in helicity-dependent 
photoproduction cross sections $(\Delta \sigma = \sigma_{1/2} - 
\sigma_{3/2})$ provide information on \cite{d1}:
\begin{itemize}
\item the spin-dependent part of the asymptotic forward amplitude
through the DHG sum rule
$$
\int^{\infty}_{\bar{\nu}}\,\frac{d \nu}{\nu}\,\Delta \sigma (\nu)\,=\,
-\frac{2 \pi^2 \alpha}{M^2}\,\kappa^2 
$$
\item the nucleon spin-dependent polarizability $\gamma$
$$
\int^{\infty}_{\bar{\nu}}\,\frac{d \nu}{\nu^3}\,
\Delta \sigma (\nu)\,=\,4 \pi^2 \gamma
$$
\end{itemize}

There are no direct mesurement of $\sigma_{1/2}$ and $\sigma_{3/2}$, 
for either the proton and the neutron. Estimates from current 
$\pi$-photoproduction multipole analyses \cite{d2}, particularly for the 
{\it proton - neutron difference}, are in good agreement with the 
relativistic 1-loop (+ $\Delta$-resonance) $\chi$PT calculations \cite{d3}
for $\gamma$ but predict large deviations from the DHG sum rule.

\vspace{0.3cm}
\begin{center}
\begin{tabular}{||c|c|c||}
\hline
 &  &  \\
 Integral  & Multipole Estimate  &  Theory \\
 &  &  \\
\hline
\hline
 DHG$_p$ $-$ DHG$_n$  &   & \\
  & $-$ 129  &  29.4   \\
(10$^{-4}$ fm$^2$) &   & \\
\hline
 $\gamma_p$ $-$ $\gamma_n$  &   & \\
  & $-$ 96  &  $-$ 104   \\
(10$^{-6}$ fm$^4$) &   & \\
\hline
\end{tabular}
\end{center}
\vspace{0.3cm}

The following two possible interpretations have been proposed:

\begin{enumerate}
\item {\it both} the higher order $\chi$PT corrections to 
$\gamma$ are large {\it and} the existing multipole are wrong
\item modifications to the DHG sum rule are required to fully 
describe the isospin structure of the nucleon
\end{enumerate}

The helicity-dependent photoreaction amplitudes, for both the 
proton and the neutron, will be measured at LEGS from the 
pion-threshold to 470 MeV. Almost 90 $\%$ of the $\gamma$ 
integral will be covered by this set of data, providing a 
reasonable comparison with the $\chi$PT predictions. This data 
will also cover about 2/3 of the DHG integral.

In these double-polarization experiments, circularly polarized 
photons from LEGS will be used with SPHICE, a new frozen-spin 
target consisting of $\vec{H}$, $\vec{D}$ in the solid phase.
Reaction channels will be identified in SASY, a large detector 
array consisting of wire chambers, scintillators and Cerenkov 
counters with a global solid angle coverage of about 80 $\%$ 
of 4$\pi$.

\section{C. Forti: Underground muons, a tool to study the
cosmic ray composition and the properties of high energy interactions}

    We discuss the importance of the detection of underground muons
  for the study of the cosmic ray composition and spectra and of the
  properties of very high energy hadronic interactions. 
    In particular, we show the application of the Monte Carlos HEMAS 
  and HEMAS-DPM to the calculation of the muon multiplicity, topology and 
  decorrelation, of the properties of multi-muon clusters and of the flux
  of muons from the decay of charmed mesons (prompt muons).

    The study of the chemical composition of cosmic rays in the energy 
 region around the knee of the spectrum is of fundamental importance for 
 the understanding of cosmic ray acceleration and propagation processes.
    The properties of high energy (TeV) muons detected deep underground are 
 strongly correlated with the mass and energy of the primary cosmic rays 
 which originated the particle shower.
    The most important features of underground muon events are :
\begin{itemize}
 \item the muon bundle multiplicity, that is the number of (almost parallel)
   muons originated by the same primary cosmic ray;
 \item  the decoherence, that is the relative distances between all pairs of
   muons which can be formed within a muon bundle;
 \item the decorrelation curve, which is the relative angle between all muon
   pairs, as a function of their relative spatial separation;
\item the number of muon clusters within the muon bundle.
\end{itemize}
    We have shown that the distribution of the bundle multiplicity 
 (called multiple muon rate) is clearly sensitive to the percentage of
 heavy nuclear species in cosmic rays. Composition models with more
 heavy nuclei predict more events with high bundle multiplicity, respect
 to lighter models.
    Also the decoherence curve is sensitive to the chemical composition,
 but is peculiar characteristic is to depend on the transverse momentum
 (Pt) distribution of hadrons (mainly pions) produced in the cascade. 
 Then, the measurement of decoherence should allow to test the hypothesis
 about the Pt distribution adopted in the model for very high energy 
 (1 - $10^5$ TeV Lab) hadron-air interactions.
    The study of decorrelation and of muon clusters are new tools recently 
 proposed to enhance the efficacity of underground measurements. 
 In particular, the number of muon cluster within a muon bundle (with 
 multiplicity larger than 8 at Gran Sasso Laboratory) is shown to be 
 sensitive both to the primary composition and to the hadronic interaction 
 properties.
    A great effort is being devoted to the comparison of different Monte
 Carlo simulations, adopting different hadronic interaction models. 
 Actually, the evaluation of the systematical uncertainties in these 
 complex calculations is one of the most difficult tasks to be accomplished.
 For example, a new Monte Carlo code (HEMAS-DPM \cite{f1}) has been realised : 
 it is based on the Dual Parton Model (DPMJET-II \cite{f2}), interfaced
 to the HEMAS shower code (containing a phenomenological interaction model
\cite{f3}). 
    The HEMAS-DPM Monte Carlo allows to simulate the interaction of hadron
 and nuclei with the air target. Also, the production of charmed particles
 is accomplished. However, calculations performed \cite{f4} indicate that the 
 detection of high energy muons generated in the decay of charmed particles 
 (the so-called prompt muons) could be a hard experimental task because the 
 ratio signal (prompt muons) to noise (ordinary muons) is less than 1$\%$
 for typical ($>$ 1 TeV) underground experiments. A challenging control of
 systematics is thus required.

\section{R. Baldini: Question marks in the nucleon time-like form factors}

Question marks related to the measurements of the Nucleon Form 
Factors are reported (for a complete discussion see \cite{bp} and
references therein), namely:
\begin{itemize}
\item the factor of 2 between "asymptotic" values of space-like and
  time-like of the proton magnetic form factor;

\item the Nucleon Form Factor mainly imaginary at 9 GeV$^2$, according
  to the FENICE results and a possible explanation within PQCD;

\item baryonium rides again (according to the $e^+e^-$ total multihadronic
  cross section near the $N {\overline N}$ threshold and according to the
  Proton Form Factor at threshold) and a short review of the old
   baryonium phenomenology;

\item the Neutron time-like Form Factors: expectations and experimental
  results.
\end{itemize}
Dispersion Relations on the log of the modulus of the Form Factors
are discussed, in order
to achieve the Nucleon Form Factors in the unphysical region.

\section{G. Pancheri: Eikonalized minijets cross-section in photon
collisions$^*$}

{\bf *}
Work in collaboration with A. Grau and Y.N. Srivastava.

A model for the parton distributions of hadrons in impact parameter space 
  has been 
constructed using soft gluon summation. This model 
incorporates the salient features of
distributions obtained from the intrinsic transverse momentum
behaviour of hadrons. Under the assumption that the intrinsic behaviour is
dominated by soft gluon emission  stimulated by the scattering process,
 the b-spectrum becomes softer and softer as the
 scattering energy increases. In minijet models for the inclusive 
  cross-sections, 
 this will counter the increase from $\sigma_{jet}$.


\begin{thebibliography}{99}
\bibitem{a1} T. Waas, N. Kaiser and W. Weise, Phys. Lett. B365 (1996) 12.
\bibitem{a2} N. Kaiser, P.B. Siegel and W. Weise, Nucl. Phys. A594 (1995) 325.
\bibitem{a3} R. Baldini et al., DEAR Collaboration, report LNF-95/055 (IR).
\bibitem{a4} K. Tanaka and A. Suzuki, Phys. Rev. C45 (1992) 2068.
\bibitem{b1} F. Abe et al., CDF Collaboration, preprint FERMILAB-PUB-96-020.
\bibitem{b2} F. Aversa et al., Phys. Lett. 210B (1988) 225;
ibid. 211B (1988) 465; Nucl. Phys. B327 (1989) 105;
Phys. Rev. Lett. 65 (1990) 401.
\bibitem{b3} S.D. Ellis, Z. Kunszt and D.E. Soper, Phys. Rev. Lett. 64
(1990) 2121.
\bibitem{b4} J. Huston, preprint CTEQ512 (1995).
\bibitem{b5} G. Curci and M. Greco, 92B (1980) 175;\\
 D. Amati et al., Nucl.
Phys. B173 (1980) 429.
\bibitem{b6} M. Greco et al., in progress.
\bibitem{LEP} The LEP Collaborations Aleph, Delphi, L3, Opal and the 
LEP Electroweak Working Group, preprint CERN-PPE/95-172.
\bibitem{Bernabeu} J. Bernabeu, A. Pich and A. Santamaria, 
\PL{B200}{88}{569}.
\bibitem{Barbieri} R. Barbieri et al., \NP{B409}{93}{105}.
\bibitem{Hollik} F. Cornet, W. Hollik and W. Mosle, \NP{B428}{94}{61}
\bibitem{Langaker} P. Langacker, preprint NSF-ITP-95-14 (1995).
\bibitem{bo} J.T. Liu and D. Ng, \PL{B342}{95}{262}; P. Bamert et al.,
preprint Mc Gill-96/04 hep-ph/9602438.
\bibitem{Ellis} J. Ellis, J.L. Lopez and D.V. Nanopoulos, preprint 
CERN-TH/95-314, hep-ph/9512288.
\bibitem{Alta} P. Chiappetta et al.,
preprint PM-96-05, hep-ph/9601306;\\ 
G. Altarelli et al., preprint CERN-TH-96-20, hep-ph/9601324.
\bibitem{d1} D. Babusci et al. (LSC Collaboration), Brookhaven report
BNL-61005 (1994).
\bibitem{d2} A.M. Sandorfi et al., Phys. Rev. D50 (1994) R6681.
\bibitem{d3} V. Bernard, N. Kaiser and U.-G. Mei{\ss}ner,
Int. J. Mod. Phys. E4 (1995) 193.
\bibitem{f1} G. Battistoni, C. Forti and J. Ranft, Astroparticle Physics, 3
(1995) 157.
\bibitem{f2} J. Ranft, Phys. Rev., D51 (1995) 64.
\bibitem{f3} C. Forti et al., Phys. Rev., D42 (1990) 3668.
\bibitem{f4} G. Battistoni et al.,
Frascati preprint LNF-95/38 (P), accepted for publication
in Astroparticle Physics (1996).
\bibitem{bp} R. Baldini and E. Pasqualucci, in Chiral Dynamics: Theory
and Experiment, A. Bernstein and B.R. Holstein eds., (1995, Springer, Berlin),
p. 312.
\end{thebibliography}
\end{document}